\documentstyle[12pt,aasms4]{article}
 
\begin{document}

\title{Neutron star magnetic field evolution,\\ crust movement and glitches} 
\author{Malvin Ruderman, Tianhua Zhu and
Kaiyou Chen}
\affil{Physics Department and Columbia Astrophysics Lab\\ Columbia University
\\538 W120th Street,New York, NY 10027}

 
\begin{abstract}
Spinning superfluid neutrons in the core of a neutron star interact strongly 
with co-existing superconducting protons. One consequence is that the 
outward(inward) motion of core superfluid neutron vortices during 
spin-down(up) of a neutron star may alter the core's magnetic field. Such core 
field changes are expected to result in movements of the stellar crust and 
changes in the star's surface magnetic field which reflect those in the core
below. Observed magnitudes and evolution of the spin-down indices of canonical 
pulsars are understood as a consequence of such surface field changes. If the 
growing crustal strains caused by the changing core magnetic field 
configuration in canonical spinning-down pulsars are relaxed by large scale 
crust-cracking events, special properties  are predicted for the resulting 
changes in spin-period. These agree with various glitch observations, 
including glitch activity, permanent shifts in spin-down rates after glitches 
in young pulsars, the  intervals between glitches, families of  glitches with 
different magnitudes in the same pulsar, the sharp drop in glitch intervals 
and magnitudes as pulsar spin-periods approach 0.7s, and the general 
absence of glitching beyond this period.
\end{abstract}


\section{Introduction}
A canonical neutron star consists mainly of superfluid neutrons, 
superconducting protons (with an abundance a few percent that of the neutrons) 
and an equal number of relativistic degenerate electrons (Fermi energy$\sim 
10^2$~Mev).  In the outer kilometer the protons clump into a lattice
of neutron-rich nuclei (the stellar ``crust'') with the neutron superfluid 
filling the space between. A spinning neutron star's superfluid neutrons 
rotate at an angular rate $\Omega$ only by establishing an array of quantized 
vortex lines parallel to the stellar spin axis, with an area density
\begin{eqnarray}
 n_V = 2m_n\Omega/\pi\hbar \sim 10^4/P{\rm (sec)}\ {\rm cm}^{-2} \ .
\end{eqnarray}
Any magnetic field which passes through
the star's superconducting protons must become very inhomogeneously structured.
In a type II superconductor, expected to be the case below the crust
and perhaps all the way down to the central core, the magnetic field becomes
organized into 
\begin{eqnarray}
 n_\Phi = B/\Phi_0 \sim 10^{19}B_{12}\ {\rm cm}^{-2}
\label{nph}
\end{eqnarray}
quantized flux tubes per unit area, with 
\begin{eqnarray}
 \Phi_0 = \pi \hbar c/e = 2\cdot 10^{-7} \ Gauss\ cm^{-2}
\end{eqnarray}
the flux in each tube.  Unlike the quasi-parallel 
neutron vortex line array, the flux tube array is expected to have a
complicated twisted structure following that of the much smoother
toroidal plus
poloidal magnetic field which existed before the transition into
superconductivity (at about $10^9$~K).
 
A spinning-down (up) neutron star's neutron superfluid vortex array must
expand (contract). Because the core of a neutron vortex and a flux tube
interact strongly as they pass through each other, the moving vortices will
push on the proton's flux tube array (\cite{sau89}, \cite{sri90},
\cite{rud91}), forcing it either
(a) to move together with the vortices, or (b) to be cut through if
the flux tube array cannot respond fast enough to partake in the
vortex motion.
Section 2 discusses possible relationships among a pulsar's $\Omega$,
$B$, and rate of change of spin ($\dot\Omega$), which discriminate between these
two behaviors. In case (a) the evolution of the magnetic field at the 
core-crust interface is well determined by the initial magnetic field 
configuration and subsequent changes in stellar $\Omega$. In case (b) the
core-crust interface field would evolve more slowly relative to changes in
$\Omega$, although qualitative features of the evolution should be similar
to those of case (a). Some  microphysics and observations, considered in 
sections 2 and 3, support case 
(a) behavior for pulsars whose spin-down (or up) ages, 
$T_s = |\Omega /2\dot\Omega |$, are not less than those of Vela-like 
radiopulsars ($T_s\sim 10^4\ years$) and case (b) behavior for the much more
rapidly spinning-down Crab-like radiopulsars ($T_s\sim 10^3\ years$)

Between the stellar core and the world outside it is a solid crust with a very
high electrical conductivity.  If the crust were absolutely rigid and a perfect
conductor then its response to changes in the core magnetic field would be
limited to rigid crust rotations.  Of course neither is the case.  
 
A high density of core flux tubes merges into a smooth field when passing
through the crust. Because of the almost rigid crust's high conductivity, it,
at least temporarily, freezes in place the
capitals of the core's flux tubes. As these flux tube capitals at the
crust-core interface are pushed by a moving core neutron vortex array, a large
stress builds up in the crust.  This stress will be relaxed when the crust is
stressed beyond its yield strength, or, if the build-up is slow enough,
by dissipation of the crustal eddy
currents which hold in place the magnetic field as it passes from the core through the
crust.  The shear modulus of a crust is well described quantitatively, but not
the maximum crust strain before yielding (and the associated yield strength).
Rough estimates have suggested a maximum yield strain, $\theta_{\rm max}$, 
between $10^{-4}$ and $10^{-3}$ (\cite{rud91}).  Nor is it known how the 
stellar crust moves when its yield strength is exceeded. By plastic flow 
(creep)?  By crumbling? By cracking?  The answer is likely to depend on the
crust temperature.  A crust's eddy current dissipation time could
be anywhere in the range $10^6 - 10^{10}$ years depending upon how the crust
was made.  A young solitary pulsar was probably born with a temperature $k_BT
\sim 10$~MeV.  As it cooled the formation of crust nuclei and their
crystallization into a crustal lattice occurred at about the same temperature,
$k_BT \sim 1$~MeV.  The impurity fraction (the probability that neighboring
nuclei have different proton numbers) has not been calculated quantitatively
and this allows a very wide latitude in the possible range for the
``impurity'' contribution to crustal resistivity.  In addition, the crust of 
an accreting neutron star spun-up to a period of a few milliseconds in a LMXB has had a very different history 
from that of a solitary spinning-down radiopulsar. The LMXB neutron star 
ultimately accretes 
more than $10^2$ times the mass of the
nuclei in its crustal lattice, mainly as He or H. Crust is continually pushed
into the core by the loading, and replaced.  As the accreted
 H and He are buried
with growing density  a series of nuclear reactions
ultimately fuse them into heavier magic number nuclei ($Z =$ 40, 50 32) 
(\cite{neg73}). This is
probably not accomplished without some explosive nuclear burning.  The
resulting reformed crust may well have an impurity fraction, electrical
conductivity, and crust thickness very different from that of a canonical 
young solitary radiopulsar.

There seems to be considerable observational, as well as theoretical, support
for the hypothesis that the surface magnetic fields of neutron stars slowly 
spun-up to become millisecond pulsars by accretion in LMXB's do indeed reflect 
the expected core field evolution at the crust-core interface (\cite{che97}, 
\cite{che93}). The core field there does appear to have had
a case (a) history : the core's magnetic flux tubes were moved in to the
spin-axis by the 
contracting neutron superfluid vortex array. Here the spin-up time scales 
($\geq 10^8\ $ years) are so very long that crustal shielding of core 
magnetic field changes is expected to be relatively easily defeated.
Rough estimates of crust properties (\cite{rud91}) indicate 
that, generally, crustal
yielding in the younger much more rapidly spinning-down pulsars 
also causes the
surface field of such neutron stars to be strongly correlated with the
configuration of the core flux which enters the crust at the core-crust
interface. (See, however, the exception for the very slowly spinning
X-ray pulsars. ) Stratification in the crust (because the $Z$ of the most 
stable nucleus varies with depth) allows mainly only two-dimensional crustal 
movement on surfaces of constant gravitational (plus centrifugal) potential.  
Where the surface field is strongest, and crustal stresses from moving 
crust-anchored core flux greatest, crustal matter would be expected to move 
with the core's moving flux, accompanied by the backflow of more weakly 
magnetized regions of the crust.  Below, except for the special
case of the very slow X-ray pulsars, we shall simply assume that
shielding by the crust of changes in the core flux emerging into it, is, at
best, temporary and unimportant even on the spin-down 
time scales of solitary radiopulsars.

In section 3 we review the expected pulsar magnetic dipole
moment evolution caused by neutron star spin-down or spun-up. It gives
young radiopulsar spin-down indices which do not disagree with observations. 
These results are not sensitive to details of just how a crust relaxes the 
growing stresses on it from the moving core magnetic flux tubes below it.
In Section 4 we consider particular consequences when that relaxation is 
accomplished by large scale crust cracking events, which cause
 pulsar timing glitches. A permanent (i.e. unhealed)
jump in spin-down rate should remain after almost all glitches. The calculated
glitch spin-period jump magnitude is
closely related to it. Both depend upon how much crust stress relaxation is
accomplished in each such cracking event. This  can be estimated very roughly
at best. However, the glitch model  does lead to predictions for the 
magnitudes of small glitches in Crab-like pulsars and of giant ones in 
Vela-like pulsars, for the intervals between such glitches, for a drop in
glitch magnitudes in long period pulsars and 
maximum pulsar period beyond which large glitches should disappear. These 
predictions are not in conflict
with glitch observations. One important consequence of the model is that
some parts of the core neutron superfluid can spin-up very slowly after
the beginning of a glitch because of the large drag in rapidly moving core
vortices embedded in a dense flux tube array. If so the canonical assumption
(\cite{alp88}) of an unobservably tight coupling between all of a core's 
neutron superfluid and the charged components of the pulsar should be 
reassessed.

\section{Core Flux Tube Movements in Pulsars}
During neutron star spin-down (e.g., in a solitary radiopulsar) or spin-up 
(e.g., by accretion in a Low Mass X-ray Binary) neutron superfluid vortices
a vector distance ${\bf r}_\perp$ from the stellar spin-axis move with a 
radial velocity
\begin{eqnarray}
{\bf v}_V = -{\bf r}_\perp \dot P/2P \ .
\label{vor}
\end{eqnarray}
As a result of this motion a force density (${\bf F}$) will build up
on the flux tube array in which these vortex lines are embedded until the
flux tubes move with, or are cut through by, the moving vortices. The core 
electron-proton plasma is almost incompressible and its abundance relative to 
the core neutrons varies with radius. Because of the extremely weak conversion 
rate for the transformations $ n\rightarrow p + e + \tilde{\nu }$ and
$ p + e \rightarrow  n + \nu $
needed to maintain a large bulk electron-proton sea transport across stellar
radii, non-dissipative motions in which the electron-proton plasma
and its embedded flux tubes move together are restricted.
We consider below mainly the alternative  where flux tubes in response
to the force on them from a changing neutron vortex array move through
the proton-electron sea with some relative velocity ${\bf v}_\phi $.
 
Magnetic field movement by eddy diffusion in an ordinary conductor is driven
by the self-stress force density of a non-force-free {\bf B}-field 
configuration:
\begin{eqnarray}
	{\bf F} = {{\bf J}\times {\bf B}\over c} \ .
\end{eqnarray}
This {\bf F} forces flux to move through the conductor with a characteristic
velocity
\begin{eqnarray}
	{\bf v}_\Phi\sim {{\bf F}c^2\over\sigma B^2}\ ,
\label{vph}
\end{eqnarray}
where $\sigma$ is the electrical conductivity of the medium. Here the force 
density ${\bf F}$ is mainly a consequence of large scale inhomogeneity in 
the field distribution,
\begin{eqnarray}
	{\bf F} = {({\bf\nabla}\times {\bf B})\times {\bf B}\over 4\pi}\ .
\label{force}
\end{eqnarray}
The time for ${\bf B}$ to be pushed out of a stationary stellar core of 
radius R would then be the usual eddy diffusion time
\begin{eqnarray}
	\tau \sim {R\over v_\Phi} \sim {4\pi\sigma R^2\over c^2} \ .
\end{eqnarray}
The resistivity $\sigma^{-1}$ in a non-superconducting degenerate 
electron-proton sea is dominated by electron-phonon scattering (\cite{bay69a}):
\begin{eqnarray}
	\sigma_{eph}^{-1} = 7\times 10^{-46}\left({10^{13} g\ cm^{-3}\over\rho_p}\right)^{3/2}T^2
\ s \ .
\label{sig1}
\end{eqnarray}
with T the temperature and $\rho_p$ the proton density. From the resistivity
of Equation~\ref{sig1} with plausible neutron star parameter and the ${\bf F}$
of Equation~\ref{force} with $|{\bf\nabla}\times {\bf B}|\sim |{\bf B}/R|$,
$\tau $ greatly exceeds $10^{10}$years. The $v_\Phi$ of Equation~\ref{vph}
would then be too small to be of interest for observable flux changes in a
spinning-down (or up) neutron star. However, when the proton sea becomes
superconducting, the $v_\Phi$ of Equation~\ref{vph} can become very much
greater. This is because of the sub-microscopic bunching of ${\bf B}$ into
the huge density of quantized flux tubes. This has two consequences. First,
a randomized electron scattering comes not only from collisions with phonons,
but also from collisions with the flux tubes themselves. The latter contributes
a much larger resistivity than that of Equation~\ref{sig1}. Second, the 
contribution to the force ${\bf F}$ that drives the flux tube motion which 
is caused by the push of moving superfluid neutron vortex-lines on flux tubes
can very greatly exceed that of Equation~\ref{force}, the self-stress 
calculated from the large scale variation of a classically smooth field.
Flux tube motion in response to some  ${\bf F}$ is possible only if the 
necessary energy dissipation accompanying it equals the work done by ${\bf F}$,
then
\begin{eqnarray}
	{\bf v}_\Phi \cdot {\bf F} = \sigma \left({{\bf v}_\Phi\cdot 
{\bf \hat{B}} n_\Phi\Phi_0\over c}\right)^2 + n_\Phi \eta v_\Phi^2,
\label{diss}
\end{eqnarray}
where the locally average ${\bf B} = n_\Phi\Phi_0 {\bf \hat{B}} $.
The first term on the RHS is the dissipation from the current flow caused
by the simultaneous motion of very many flux tubes (It has typically been 
neglected in the literature. Its importance was emphasized by P. Goldreich 
(1993).). 
In writing Equation~\ref{diss} we make the implicit assumption that
the original array of flux tubes moves but no new flux loops are created or
existing ones reconnected and destroyed. They may not
be valid except in the limit of very small $v_\Phi$.
The conductivity $\sigma $ is that for (electron) current flow in the 
${\bf E} = {\bf v}_\Phi \times {\bf B}/c$ direction, i.e. perpendicular to
${\bf B}$. For a given ${\bf B}$ this contribution to
dissipation is not sensitive to details of flux tube radii or the 
magnitude $\Phi_0$ except through the dependence of $\sigma$ upon both of 
them. 

The second term on the RHS is from the direct drag force (along ${\bf v}_\Phi$)
on individual flux tubes pushing through the electron sea. The drag coefficient
(force per unit length of flux tube $ =\eta v_\Phi$) on an isolated solitary
flux tube (\cite{jon87}, \cite{har86}), 
\begin{eqnarray}
	\eta ={{3\pi\Phi_0^2 e^2n_e}\over{64 \Lambda_* c E_f}},
\label{drag}
\end{eqnarray}
with $E_f$ the electron sea Fermi energy and $\Lambda_*$ the radius of a
flux tube ($\sim 10^{-11}\ cm$). [$\Lambda_* = 
(m_pm_p^*c^2/{4\pi e^2\rho_p})^{1/2}$ with $m_p^*$ the effective proton mass and $\rho_p$ the proton plasma density.] 

The electron resistivity, $\sigma^{-1}$, now has two contributions. One
is the contribution from electron-phonon scattering of Equation~\ref{sig1};
the other is from scattering of electrons on the flux tubes themselves. Because
the magnetic flux is bundled into intensely magnetized flux tubes at each of 
which electrons are scattered through a finite angle ($\Delta\Phi$), there is
a drag along the electron velocity proportional to $(\Delta\Phi)^2$ at each scattering. (Equivalently the circular  trajectory of an electron in a ``uniform''
{\bf B} is replaced by a polygon with a random scattering component 
$\sim [\overline{(\Delta\Phi)^2}]^{1/2}$ at each vertex.) Because the 
separation between scatters 
($\gg (\Phi_0/B)^{1/2}\sim 3\times 10^{-10} B_{12}^{-1} cm$) is very large
 compared to
$\hbar c/E_f\sim 10^{-13} cm $, there is negligible
interference between scattering at different vertices.)
The drag along the electron velocity is just that from Equation~\ref{drag}.
It contributes a resistivity
\begin{eqnarray}
	\sigma_{e\Phi }^{-1} = {\eta n_\Phi\over e^2n_e^2},
\label{sig2}
\end{eqnarray}
with $n_e$ the number density of electrons. The contribution of 
Equation~\ref{sig2} to 
\begin{eqnarray}
	\sigma^{-1} = \sigma_{e\Phi}^{-1} + \sigma_{eph}^{-1}
\end{eqnarray}
is generally much more important than that of Equation~\ref{sig1}. (
For typical neutron star parameters $\rho_p\sim 10^{13}\ g\ cm^{-3}$ and
$T = 10^8 K$, $\sigma_{eph}^{-1}\sim 10^{-29}$ s while $\sigma_{e\Phi}^{-1}\sim
10^{-27} B_{12}$.) If we neglect it we can approximate a very small flux tube
velocity in the direction of a ${\bf F}$ perpendicular to ${\bf B}$ by the
exact analogue of Equation~\ref{vph}
\begin{eqnarray}
	{\bf v}_\Phi\sim {{\bf F} c^2 \over \sigma n_\Phi^2 \Phi_0^2},
\label{vphi}
\end{eqnarray}
with an effective conductivity
\begin{eqnarray}
	\sigma = \left({{e^2 n_e^2}\over {\eta }} + {{c^2\eta} \over
	{\Phi_0^2} }\right) n_\Phi^{-1}\ .
\label{cond}
\end{eqnarray}
We note that ${\bf v}_\Phi\rightarrow 0$ when $\eta\rightarrow 0$ because of
infinite electron conductivity, and also when $\eta\rightarrow \infty $ because
of the infinite drag on a solitary moving ( with respect to the e - p sea ) 
flux tube. The contribution of the second term on the RHS of Equation~\ref{cond} to $\sigma $ is generally negligible in typical pulsars.

To evaluate the maximum $|{\bf v}_\Phi|$ before the cutting through of a flux 
tube array by a moving vortex array we must now consider the maximum ${\bf F}$
just before cutting through begins. From Appendix A, this is, roughly,
\begin{eqnarray}
	F_{max} \simeq {\pi n_V\over 8} B_\Phi B_V\Lambda_*\ln\left({\Lambda_*
\over\xi}\right),
\label{fmax} 
\end{eqnarray}
with $n_V$ the vortex area density of Equation~\ref{nph}, 
$B_\Phi\sim \Phi_0/\pi\Lambda_*^2$ the magnetic 
field within a flux tube, $B_V\sim B_{\Phi}$ the magnetic field within a 
vortex line, and $\xi (< \Lambda_*)$ the BCS correlation length of the Cooper
pairs in the superconducting proton sea. [ The force density of 
Equation~\ref{fmax} greatly exceeds that from flux line curvature 
(\cite{har86}) or flux tube buoyancy (\cite{mus85}).] From 
Equations~\ref{vphi}, \ref{cond}, and \ref{fmax}
the maximum velocity ($v_c$) with which a moving vortex array can push 
a flux tube array through the electron-proton sea in which it is embedded would
be
\begin{eqnarray}
        v_c = \beta \left({\Omega\over 100}\right)
\left({10^{12}G\over B}\right) 10^{-6}\ cm\ s^{-1} \ ,
\label{Vcr}
\end{eqnarray}
i.e. $v_c$ is proportional to the ratio of vortex line density to flux tube
density.
The proportionality constant, $\beta$, is independent of $\Omega $ and $B$ but does depend upon
properties of neutron star matter below the crust: 
\begin{eqnarray}
\beta = 0.4\times \ln\left({\Lambda_*\over\xi}\right)
\left({B_V\over 10^{15}G}\right)
\left({B_\Phi\over 10^{15}G}\right)\left({60Mev\over E_f}\right)
\left({10^{36}cm^{-3}\over n_e}\right)\ .
\label{bet}
\end{eqnarray}

The constant $\beta$ depends upon imprecise estimates of the
vortex flux-tube interaction, the flux-tube spacing along moving vortex lines,
the angle between local ${\bf B}$ and ${\bf\Omega }$, etc. 
However the main problem with applying Equations~\ref{Vcr} and \ref{bet}
to flux
tube motion may be the implicit assumption that ${\bf v}_\Phi$ is so small that $n_\Phi$ (and thus local ${\bf B}$)
in it is qualitatively unaffected by the electric currents induced
by the flux tube motion, i.e. that the effect of ${\bf F}$ is only to move
the preexisting flux tubes which remain locally straight and uniformly 
distributed. Further, the geometrical distribution and
motion of flux tubes may, in reality, be quite complicated with flux tubes, 
the electron-proton seas, and neutron vortex lines moving together without 
cutting-through
in many regions and with vortices cutting through flux tubes in others. 
We emphasize that for two dimensional motions of the electron-proton sea
in the spherical layer just below the crust ( the only core layer which 
directly affects the surface field ) stratification does not restrict flux 
tube crowns in the most magnetized regions from being moved by vortex push
 from initial positions near the spin-axis all the way down to the equator
during spin-down. We
shall, therefore, consider Equation~\ref{Vcr} as a 
phenomenological one for the behavior of magnetic flux tubes in the stellar
core layer just below the crust-core interface 
with $B$ the pulsar dipole field strength inferred
from spin-down. We take $\beta\sim 1$, about the value expected from 
Equation~\ref{bet}, but even more 
because Equation~\ref{Vcr} then leads to a good 
description of various observed properties of young spinning-down radiopulsars.

The velocity $v_V$ as a function of $r_\bot$ and $v_c$ of 
Equation~\ref{Vcr} with $\beta =1$ is sketched in Figure 1 for a 
Vela-like pulsar with $\Omega \simeq 100 s^{-1}$, and  $B= 10^{12}$ G. For 
$r_\bot <r_c$ the neutron superfluid vortex expansion velocity (proportional 
to $r_\bot$) is slow enough to carry all flux tubes with the expanding vortex 
array, at least in the core layer just below
the crust; flux tube cut through occurs  for $r_\bot >r_c$.
From Equations~\ref{vor} and \ref{Vcr}
\begin{eqnarray}
	r_c \simeq \left({T_s\over 10^4yrs}\right)
	\left({\Omega_2 \over B_{12}}\right) 10^6 cm,
\label{r_cr}
\end{eqnarray}
with $T_s$ the pulsar spin-down time scale (age). Then for $T_s\Omega_2/B_{12}
\geq 10^4 yrs$, i.e. for Vela-like pulsars and those much older, $r_c \geq 10^6 cm$,
i.e $r_c\geq $ the stellar radius R and all flux would move out with the 
${\bf v}_V$ of the vortex array. For Crab-like pulsars with $T_s$ an
order of magnitude smaller than that for the Vela pulsar most of the flux 
array (except that within 
$r_\bot\sim 10^{-1}R$ of the spin-axis) would move out much more slowly than 
the neutron vortices.
As indicated in Figure 1, however, it is not yet known how fast that 
cut-through flux tube outward flow 
should be.

\section{Surface magnetic field evolution and spin-down indices}
Based upon the above assumptions and estimates about the interaction between 
a pulsar core's arrays of superfluid neutron vortices and superconducting 
proton flux tubes, we consider below consequences of a greatly simplified
model for the evolution of magnetic fields in spinning-down pulsars:
\begin{enumerate}
	\item The crust and core magnetic fields will be described as if 
they were axially symmetric around
the spin axis (clearly in contradiction to what is required for a pulsar's
rotating radio beams). The important consequence is that core flux tubes 
can then move outward only by pushing through the core's electron-proton sea, even if their actual motion is more complicated (and
might not involve such push through in many regions).
	\item When $r<r_c$ of Equation~\ref{r_cr} with $\beta = 1$,
flux tubes move outward 
with the velocity ${\bf v}_V$ of Equation~\ref{vor}.
	\item When $r>r_c$ flux tubes are moved outward with the smaller 
velocity $v_c$ of Equation~\ref{Vcr}. For example in the Vela pulsar
${\bf v}_\Phi \simeq {\bf v}_V$ for almost all flux tubes, but in the
Crab pulsar most flux tubes would  not keep up with the core's neutron
vortices. Rather,
\begin{eqnarray}
{\bf v}_\Phi (Crab) \sim {\bf v}_V (Vela)\ .
\end{eqnarray}
	\item The surface fields of the neutron star reflect those of the
core at the core-crust interface. ( This, probably, would not be 
accomplished for exact axial symmetry. In a more realistic model it would be
expected only for the most strongly magnetized regions since some crustal
backflow (where $B$ is weakest) would be expected to allow the strongly
forced crust movement where $B$ is largest.)
\end{enumerate}
We consider next a comparison of the predictions of such a model to 
observations of $\ddot P, \dot P$, and P for some of the younger pulsars.

In this model the core and surface magnetic field configurations of a neutron 
star 
depend not only on the star's spin history, but also on its (quite unknown) 
initial field configuration.  It is often convenient in calculations to
assume the surface field to be that of a central dipole but there are no 
physical arguments supporting this special configuration as there is , for 
example, for the earth's surface field where the surface is very far from the 
core dynamo currents. More plausible might be some (random) mixture of
higher moments (\cite{bar82}), or a strongly off-center dipole from a 
toroidal field (originally amplified by initial differential rotation) which 
has pushed out through the stellar surface in some region.  An initial 
``sunspot-like'' surface field configuration seems needed to describe the 
evolution of some neutron stars which are spun-up to become very fast 
millisecond pulsars (\cite{che93}): most of the magnetic flux from
each of these stars spin-hemispheres returns to the star in the same hemisphere
as that from which it originates.  

With an  axially symmetric magnetic
field configuration the spin-down rate of a solitary neutron
star depends almost entirely on its net dipole moment $({\bf \mu})$ which 
can vary and its moment of inertia. The expected evolution of such a dipole 
moment is shown in Figure 2 together with 
inferred moments (from observed spin-down rates) of radiopulsars.  Three 
common evolutionary stages are predicted for all pulsars: 
\begin{description}
	\item[Stage a - b)] In young Crab-like pulsars, $r_c$ is much smaller
than the $10^6$cm stellar radius. In most of the core 
$r_\bot > r_c$. Superfluid vortices there cut through magnetic
flux tubes and $|{\bf v}_\Phi | < |{\bf v}_V|$. Because 
$\dot \Omega \propto \mu^2\Omega^3/Ic^2$
(essentially from dimensional arguments) with I the star's moment of inertia,
 the spin-down index
\begin{eqnarray}
n\equiv {\ddot \Omega\Omega\over \dot \Omega^2} = 3 -
T_s \left( {4\dot \mu \over \mu} - 2 {\dot I\over I} \right),
\label{index}
\end{eqnarray}
Measured values of $n$ are given in Table 1.
\begin{table}[tb]
\centering
\begin{tabular}{|c@{\hspace{0.5cm}}|c@{\hspace{0.5cm}}|c@{\hspace{0.5cm}}|c@{\hspace{0.5cm}}|c@{\hspace{0.5cm}}|} 
\multicolumn{5}{c}{TABLE 1.}\\
\multicolumn{5}{c}{Pulsar spin-down indices} \\
\multicolumn{5}{c}{}\\ \hline\hline
PSR  & $T_s$(yr) & n & $n_{Model}$ & Ref. \\ \hline
Crab & 1300 & 2.5 & 2.6 &\cite{lyn88} \\
1509-58 & 1500 & 2.8 & 3 & \cite{kas94} \\
0540-69 & 1700 & 2.0 & 2.7 & \cite{man89} \\
Vela & 11000 & 1.4 & 2  & \cite{lyn96} \\ \hline
\end{tabular}
\end{table}
Plausible $\dot I/I$ (\cite{alp96}) seem too small to be a promising 
explanation of the large 
$3 - n$ of Vela, and we neglect its contribution to 
Equation~(\ref{index}). The model of Section 2 suggests 
\begin{eqnarray}
|\dot\mu/\mu | \sim {|{\bf v}_\Phi|\over |{\bf v}_V|}(4T_s)^{-1}
\end{eqnarray}
with $\dot\mu/\mu > 0$ for a ``sunspot''-like field configuration,
as long as magnetic flux has not yet been pushed out of the core at
 the (spin) equator. 
Then, for such (shorter period) pulsars
\begin{eqnarray}
3-n\sim v_\Phi/v_V \ .
\label{index2}
\end{eqnarray}
Insofar as $r_c > R$ in Vela, $v_\Phi = v_V$ for that pulsar. With
this approximation
the model predicts n=2 for Vela. In the more general case the assumption
$\dot\Omega\propto \mu^2\Omega^3$ is replaced by $\dot\Omega\propto 
(\alpha\mu_\bot^2 + \beta\mu_\parallel^2)\Omega^3$ where $\mu_\bot$ is the 
component of $\mu$ perpendicular to ${\bf\Omega}$ and $\mu_\parallel$ is the
parallel component. For time independent $\alpha$ and $\beta$
\begin{eqnarray}
	n = 3 + \left(\beta\mu_\parallel^2 + {2\beta\mu_\parallel\dot\mu_\parallel\Omega\over \dot\Omega (\alpha\mu_\bot^2 + \beta\mu_\parallel^2)} - 1\right){ v_\Phi\over v_V}
\end{eqnarray}
For a spinning dipole in a vacuum, $\beta = 0$ and equation~\ref{index2} is 
recovered with n=2 for Vela.
For much more rapidly spinning 
Crab-like
pulsars with much smaller spin-down ages, but with $v_\Phi$ still the same as
that of Vela because of the cut-through of their magnetic flux tubes by 
their more rapidly expanding vortex-arrays, the model gives 
\begin{eqnarray}
 3 - n = \left( 3 - n\right)_{vela}\left({\Omega T_s\over B}\right)
\left({\Omega T_s\over B}\right)_{vela}^{-1} \ .
\label{vphiV}
\end{eqnarray}
Equation~\ref{vphiV} is used to give the other spin-down indices in the 
$n_{model}$ column of Table 1. 
Comparisons with observations are satisfactory except for
PSR 0540-69. However, it has been suggested (\cite{oge90}) that the
braking index of PSR 0540-69 could be 2.7 instead of 2.0 because of a glitch
just before their period measurements of this pulsar. If this is indeed the
 case
the agreement would be satisfactory here  also. 
For pulsars older than $10^4$ years but not very much older, flux tubes are 
predicted to move outward  with the
same velocity as vortices. For them $|{\bf v}_\Phi | \sim |{\bf v}_V |$ and
$n \sim 2$. [If ``magnetars'' (\cite{tho93}), pulsars born with huge 
$(B \sim 10^{15}\ {\rm G})$ magnetic fields, exist 
they would spin-down so rapidly ($P\sim 10$~s after $10^4$ yrs) that $v_\Phi 
\ll v_V$. Then for most of their early lives $n\sim 3$ and $\mu$ would not be 
much diminished by the spin-down.]

\item[Stage b - c)] Until an age $T_s \sim 10^4$~years is exceeded, movement 
of the most strongly magnetized surface patches toward the spin equator is 
predicted to be much slower than that of the core's neutron vortex lines.  In 
much older pulsars, with
flux tubes and vortices moving together, a significant fraction of the flux
should begin to reach the spin-equator and be pushed out through the crust-core
interface region into the deep crust.  Subsequently, the core's vortex
array no longer controls the movement of that flux.  The  movement  of a
typical flux tube is sketched in Figure 3 (for an initial non-sunspot
configuration).  When enough flux is expelled from the core, the huge
stresses that build up in the crust (whose rigidity alone  prevents rapid
reconnection between north and south polar regions of core ejected
flux) can become large enough to  exceed the yield strength of the crust.  
Then reconnection allowed by
crust breaking and Eddy dissipation begin. [ The magnetic stress on the 
crust could reach or even exceed
$BB_c/8\pi$, with $B_c \geq 10^{15}\ {\rm G}$ the magnetic field within a 
flux tube. The yield strength of a neutron star's crust when stressed over
a surface area of radius$\sim$R is $\mu\theta_{max}\Delta/R$ where $\mu$ is 
the deep crust shear modulus, $\Delta$ the crust thickness, and $\theta_{max}$
the maximum strain before yielding by breaking or plastic flow. (This crust
strength is about $10^{-1}$ the ``yield stress'' of crustal matter.) Because 
$\theta_{max}$ depends upon uncalculated details of crustal dislocations and
impurities, its value is uncertain. Typical estimates for it give 
$\theta_{max} < 10^{-3}$. Then $BB_c/8\pi\sim 10^{26} dyne\ cm^{-2}\geq \mu\theta_{max}\Delta /R\sim 10^{25} dyne\ cm^{-2}$. In addition, and perhaps of
greater significance the time scale for reconnection because of Eddy 
diffusion through the thin crust is diminished because of the special 
core-expelled magnetic field geometry : radial field B is much smaller than
 tangetial field $B_c$. The relevant Eddy diffusion time $\sim (4\pi\Delta^2/c^2)\times (crust \ conductivity)$. The unknown impurity contribution to 
crust conductivity makes quantitative estimates of the diffusion time 
quite uncertain. It is not implausible that it can be less than the $10^6$ year lifetime of most radiopulsars. ]  The surface field evolution of a spinning-down star 
after most north and south pole regions reach the core's spin-equator and
ultimately reconnect is sketched in
Figure 4. The unreconnected flux still left in the stellar core is roughly
proportional to $\Omega$.  Then $\mu \propto \Omega$ and 
Equation~(\ref{index}) 
gives $n=5$. This predicted decline with increasing spin-period 
$P$ in the dipole component of the 
surface field is shown as segment (b - c - d) in Figure 2.  We see no reason 
for those strongly magnetized
north and south polar surface regions (magnetized ``platelets'') which have 
been pushed to the spin-equator after  some fixed time 
to contain exactly equal amounts of flux. Any excess in the
equatorial zone not canceled by reconnection would be connected to some other
magnetized region which has not yet reached that zone (e.g., because it started much
closer to the spin axis and, therefore, has moved away from it much more
slowly).  This is sketched as the region N$^\prime$ in Figure 4.  The direction
of the remaining dipole ${\bf \mu}$ depends on details of the initial
configuration; only its diminished magnitude is a robust prediction.

Observations are not in conflict with the model curve segment b - c of 
Figure 2. We note especially the eight $10^4$~year old radio-pulsars still in 
supernova remnants. Unless strong $\mu$ reduction does indeed begin , similar 
to that indicated as segment b - c, there is a puzzle in trying to understand 
the Figure 2 data.  Where will the descendants of these 8 Vela-like pulsars in 
SNR's be observed ? If $\mu$ is constant the number of pulsars in any 
fractional period interval $\Delta P/P$ should be proportional to $P^{-2}$.  
Thus there should then be of order $10^3$ pulsars with $P\sim 1$~s with a 
dipole moment similar to that of these 8 Vela-like pulsars. Where are they? 
The total number of slower pulsars
actually observed does not particularly contradict this expectation but their inferred $\mu$ is clearly diminished. With the observed $n\sim 1.4$ 
in Vela, this absence of a very large number of descendants of Vela-like 
pulsars with the same $\mu$ as that of Vela or even a greater one 
would be even more dramatic.

\item[Stage c - d)] Most radio-pulsars  die before their spin-periods 
exceed several
seconds.  However, some will be in binaries where interaction with a
companion (via winds, accretion disks, common envelopes) may spin the neutron
stars down to very much greater periods.  The core magnetic  field would
continue to drop, but ultimately a lower limit would be reached where a crust's
strength and high conductivity freezes the crust field even after almost all
flux has been expelled from the core.  Because of quantitative uncertainties
about the crust's yield strength it is not known just when this will occur. 
Segment (d) in Figure 2, where crust flux freezing is assumed to become
 effective, is, 
therefore, mostly a plausible guess.  The magnetic moments of slow X-ray 
pulsars should retain such a value until crustal eddy currents decay even
though for some of them $P\sim 10^3$~s.  One characteristic of the surface
field of such spun-down pulsars should reflect the special way in which their 
dipole field was diminished.  Initially separated strongly magnetized 
``platelets''  were first pulled
away from each other and, if they had opposite polarity, later had their fields
reconnected after they reach the spin-equatorial zone.  However, each strongly
magnetized platelet is much less likely to become stressed in a way which would
have caused it to fragment:  wherever significant field remains on the surface
of a spun-down pulsar it should still tend to have the same strong value 
that much of the entire
stellar surface had originally.  Consequently, in slowly spinning pulsars,
polar cap magnetic fields measured by cyclotron resonance features in X-ray
spectra should give a very considerably higher magnetic field strength than
that inferred from observations which are sensitive only to the stellar
magnetic dipole moment (e.g., $(P\dot P)^{1/2}$ in radio-pulsars and X-ray pulsars). 
This may already be implied in observations of the accreting binary which
contains the $P=1.2$~s X-ray pulsar Her X-1.  Its X-ray cyclotron resonance
feature gives $B\sim 5\cdot 10^{12}\ {\rm G}$ (\cite{tru78}), but accretion 
disk modeling is best fit for a dipole $B \leq 10^{12}$~G (\cite{gho79}). 
Stages {\bf def} and {\bf deg} for spun-up pulsars and their relation to 
millisecond pulsar observations have been discussed 
elsewhere (\cite{che93}, \cite{che97}).
\end{description}

\section{Glitches}
The surface magnetic field evolution in the pulsars considered above is not
sensitive to details of the associated crust movements.  For the warm crusts
of very young radiopulsars most of the crustal stress from spin-down induced
motion of core-flux should be relaxed by plastic flow (``creep'').  For cooler crusts,
this is no longer expected to be the case.  The transition to a more brittle
crust response has been estimated to be at temperatures of a few $
10^8$~K (\cite{rud91}), about that in the deep crustal layers of $10^3$~year 
old pulsars like the Crab.  In cooler spinning-down neutron stars the forced 
movement of the most strongly magnetized surface patches may be accomplished 
by large scale crust cracking. The sudden crustal movement might itself 
be the cause of
crustal neutron superfluid vortex line unpinning or it might trigger a hydrodynamically supported unpinning avalanche (\cite{alp93}). Either
 would cause sudden changes in the 
stellar spin-period which suggest various features of 
observed spin-period ``glitches'', but they seem to differ in their predictions about
permanent changes in spin-down rates.

Figure 5 shows the magnitudes of the 34 glitches (sudden fractional jumps in
pulsar  spin frequency $\Omega$) reported by Lyne, Pritchard and Shemer (1995)
vs. the spin-down 
age $|\Omega/ 2\dot\Omega |$ of the glitching pulsars. Figure 6 shows their 
estimated ``glitch activity'' (the sum of all detected $\Delta \Omega/\Omega$ 
devided by the total observation time) as a function of pulsar spin-down 
age .
These observed glitch activity rates support the proposal 
(\cite{and75}, \cite{alp77}, \cite{alp84}, \cite{alp93}, \cite{rud76}) 
that
the main cause of the jumps in pulsar spin rate in a glitch is a sudden
spin-down of the crust's inter-nuclear neutron superfluid.  Because that
superfluid's vortex lines can be strongly pinned to the lattice of crust
nuclei, the crust neutron superfluid may not spin-down smoothly with the rest
of the star.  If crust neutron vortex lines move outward from the spin-axis
only in discrete events (glitches), sudden spin-up glitches will be observed 
for the rest
of the star.  If these pinned vortices do not move from their pinning 
sites between glitches, the part of the crust superfluid neutron 
angular momentum ($\Delta J_{csf}$) which is not diminished  during the 
spin-down intervals between glitches ($\tau_g$) is
\begin{eqnarray}
	\Delta J_{csf} = I_{csf}\dot\Omega \tau_g \ .
\label{ang1}
\end{eqnarray}
$I_{csf}$ is the 
moment of inertia of the crustal superfluid neutrons whose spin is determined
by those vortex lines  which do not unpin between glitches. 
During one or after many glitches the drop $\Delta J_{csf}$ is accomplished
and balanced by spin-up of the
other parts
of the neutron star. Then the glitch activity  is 
\begin{eqnarray}
 {\Delta \Omega\over \Omega} \cdot {1\over  \tau_g} \sim {I_{csf}\over I_*}
{\dot\Omega\over \Omega},
\label{gact}
\end{eqnarray}
where $\Delta \Omega / \Omega$ is the observed glitch magnitude,
$I_* - I_{csf}$ ( $I_{*} \gg I_{csf}$) is the moment 
of inertia of all the parts of the star which, before a spin-period glitch 
is resolved, share that angular momentum increase which balances the sudden 
glitch associated decrease in that of crust neutron superfluid. Table 2 gives 
the model result of 
Equation~(\ref{gact}) for $I_{sfc}\simeq 1.5\times 10^{-2}I_*$ 
(a typical value of
the moment of inertia of crustal neutron superfluid from neutron
star models) with the glitch activity rates of those young pulsars 
which have been observed to glitch more than once and thus
allow an estimate  of their glitch activity. The comparison between 
Equation~\ref{gact} and observations is also shown in Figure 6. The agreement 
with Equation~\ref{gact} is satisfactory except for the young 
Crab family. The cause of this discrepancy will be discussed below. 

A quantitative calculation of $I_*$ is complicated because the core's neutron 
superfluid vortices are immersed in and push on the core's flux tube array. 
All of the core neutron superfluid vortices would not be able to move inward 
quickly in response to the sudden glitch associated spin-up of the core's 
\begin{table}[tb]
\centering
\begin{tabular}{|c@{\hspace{0.5cm}}|c@{\hspace{0.5cm}}|c@{\hspace{0.5cm}}|c@{\hspace{0.5cm}}|c@{\hspace{0.5cm}}|}
\multicolumn{5}{c}{TABLE 2}\\
\multicolumn{5}{c}{Pulsar activity in Young Pulsars} \\
\multicolumn{5}{c}{}\\ \hline\hline
& Age & Post-glitch healing fraction & \multicolumn{2}{c|}{Glitch activity($10^{-7} yr^{-1}$)} \\ \cline{4-5}
PSR     & $\log$(age (yr)) & for $\Delta\Omega/\Omega$ & Observed & Equation~\ref{gact}.\\ \hline
0531+21 & 3.10 & 80\%  & 0.1             & 62    \\
1509-58 & 3.19 &  ?   & $\sim 0$         & 51  \\
0540-69 & 3.22 &  ?   & ?                & 47  \\
0833-45 & 4.05 & 13\%  & 7                & 7     \\ 
1338-62 & 4.08 & 1.1\% & 7                & 7     \\
1800-21 & 4.20 & 7\%   &  ?               & 5     \\
1706-44 & 4.24 & 11\%  &  ?               & 4  \\
1737-30 & 4.31 & 3\%   & 4                & 4  \\
1823-13 & 4.33 & 7\%   & 4                & 4 \\
1727-33 & 4.41 & 4\%   &  ?               & 3 \\
1758-23 & 4.77 & 0.1\% & 1                & 1 \\ \hline\hline
\multicolumn{5}{c}{ Note: All data are taken from Shemar and Lyne(1996)}
\end{tabular}
\end{table}
electron-proton plasma (tied to the crust lattice by the strong internal 
magnetic field) (\cite{din93}). It would
not include the core neutron superfluid whose vortex lines would have to 
push flux tubes through the electron-proton sea or to cut through their 
surrounding flux tubes in a time too short to be observed in a glitch.
$I_*$ would then be very significantly less than the total moment of inertia 
of the star. The straight line in Figure 6, Equation~(\ref{gact}) 
with $I_{csf}/I_* = 1.5\times 10^{-2}$, fits observations except for the very young 
Crab-like family and the oldest pulsars $(T_s > 3\cdot 10^6$~years). If $I_*$ 
were to equal the total stellar moment, this ratio gives a relatively large 
$I_{csf}$ implying a stiff core equation of state to give a thick 
enough crust. On the contrary, an important softening may be a consequence of 
a K-meson condensate (\cite{bro94}).
In the absence of a quantitative calculation of $I_*/I$, which would probably
also need detailed knowledge of the core's flux tube array to support a 
calculation of the time history for core neutron vortex response, it may be 
premature to
draw quantitative conclusions about neutron star structure from fits of
$I_{csf}/I_*$ to pulsar glitch data.

Equation~(\ref{gact}) is not a unique consequence of any one among various 
glitch
theories based upon the discontinuous spin-down of crust neutron superfluid.
It holds, for example, as long as  each crust cracking event shakes free 
only some fraction
of the crust neutron superfluid's pinned vortex lines so that a typical 
pinned vortex line survives several glitches before it is ultimately
unpinned (or even if there is no glitch vortex unpinning but only a
shift in their position because of a sudden movement of the pinning sites 
(\cite{rud76})). It would also hold if the repeated crust neutron vortex 
unpinning events have a purely hydrodynamic origin and development 
(\cite{alp93}), and may well remain valid for other kinds of glitch models 
(\cite{lin96}). There are, however, other glitch observations
which may discriminate among glitch models, in particular, those which 
are based only on spin-up vs.\ those which also have glitch associated crust 
breaking displacements.  

We consider below the interpretation of glitch features within the 
framework of the crust cracking model in which some relaxation of the
crustal stresses from core flux tube movement is the prime cause of a glitch.
\begin{description}
\item[a)] {\it The Crab pulsar's dipole magnetic field appears to jump in each
major Crab glitch.} The glitch history of the Crab pulsar is shown in Figure 7 
for spin-rate changes relative to a prediction extrapolated from initial 
observations for $P$, $\dot P$, and $\ddot P$. After
each of the two major glitches there is a permanent change in $\dot P$
indicating a crust spin-up rate change $\Delta \dot \Omega/\dot \Omega \sim
4\cdot 10^{-4}$. Each repeated  $\Delta\dot \Omega$ is much too large to be 
understood as coming from a plausible sudden shape change. There are two 
much more credible interpretations for the $\dot \Omega$ jumps: the spin-down 
torque might have suddenly increased in the glitch, or the effective crustal 
neutron superfluid's spin-down  moment of inertia might have decreased
because of some rearrangement of crustal vortex pinning (\cite{alp96}). 
This jump is a relatively huge effect; it can be seen to be very much greater 
than the relatively tiny $\Delta\Omega/\Omega$ of the glitch (most of which is 
also quickly healed). The first  explanation is a natural and
necessary consequence of local crust cracking causing a sudden movement of a
strongly magnetized platelet. We note that the sign of $\Delta \dot\Omega$ 
would then imply a sudden, unhealed
increase in the dipole moment for each major Crab glitch; this is  consistent
with the sign of $\dot \mu$ for more gradual changes inferred 
from the Crab spin-down index (Table 1).
The presumed fractional dipole increase corresponds, roughly, to a sudden
magnetized surface patch displacement (toward the equator) of $\Delta s
\sim 2\cdot 10^{-4}\>R$. This $\Delta s$ does not seem implausible when 
compared with rough estimates of  how large a healing crack displacement (if 
any) could be expected when the crustal yield strength is exceeded (
a $\Delta s/R$ somewhat less than the maximum yield strain).  We  assume 
below that this $\Delta s$ (and the associated $\Delta\dot\Omega/\dot \Omega$) 
value is  common to
all major glitches in rapidly spinning pulsars since it depends only on the
properties of a pulsar's crust, not on its period, magnetic field, or
spin-history.  Unfortunately, it is difficult to know from present data if this
is the case.  It is, however, not inconsistent with Vela pulsar glitch data 
(cf. {\bf b)}). 
	\item[b)]{\it The glitch interval for the Vela pulsar is 3 years.} 
According to Equation~\ref{vor} strongly magnetized platelets on Vela's crust 
should move toward the spin equator at an angular rate $\sim T_s^{-1}$. If 
this is accomplished by repeated crust breaking glitch events a time $\tau_g$ 
apart, then $\tau_g \sim (\Delta s/R) T_s \sim 2$~yr. This is  close to what 
is observed for Vela.  The related question of whether there is an unhealed
$\Delta\dot \Omega/\dot \Omega \sim 4\cdot 10^{-4}$ in Vela after each
glitch is not answered directly because, in distinction to Crab glitches,
a new Vela glitch occurs before healing from the previous glitch is
complete enough.  However, Vela's observed 1.4 spin-down index
could be interpreted solely as the consequence of an unhealed
$\Delta\dot \Omega/\dot \Omega = (3-n)/2 \tau_g T_s^{-1} \sim 0.8
\tau_g T_s^{-1} \sim 2\cdot 10^{-4}$ after each glitch, i.e., the near
100\% growth in magnetic moment during a spin-down time implied by $n=1.4$
might indeed be accomplished in discrete jumps at glitches.
This is not the case, however, for Crab glitches which are too infrequent to
contribute significantly to the Crab's $3-n \sim 0.5$. We note that in the 
Vela-like group it would also follow from Equation~\ref{gact} that such 
glitches have a magnitude
\begin{eqnarray}
	{\Delta\Omega \over\Omega}\sim {\tau_g\over 2T_s}\cdot 10^{-2}
	\sim 10^{-6},
\end{eqnarray}
near what is observed.
\item[c)] {\it The major Crab glitches are only a few times $10^{-2}$
as strong as the giant ones in the older pulsars. Glitches have not
been seen at all in PSR's 1509-58 and 0540-69.} The defining characteristic of 
a glitch is the  jump in the spin-rate of the pulsar crust presumed to be
caused by the sudden  small spin-down of some crustal neutron superfluid. 
The crust is a layered structure. The deep crust where such 
vortex pinning is relevant consists of three layers, some of whose
physical properties are estimated in Table 3. The nuclear charge of the 
most stable nucleus (Z) and the number density of nuclei ($n_Z$) are taken 
from the calculations of Negele and Vautherin (1973). In the deep crust
these nuclei form a coulomb lattice (i.e. the electron sea has a negligible 
polarization). The crustal lattice melting temperature ($T_m$) is then 
well approximated by $k_BT_m\sim (Ze)^2n_Z^{1/3}/180$. The $T_b$ column of
Table 3 is $10^{-1}$ the calculated crust lattice 
melting temperature.
\begin{table}
\centering
\begin{tabular}{|c@{\hspace{1cm}}|c@{\hspace{1cm}}|c@{\hspace{1cm}}|c@{\hspace{1cm}}|} 
\multicolumn{4}{c}{TABLE 3.}\\
\multicolumn{4}{c}{Properties of Deep Crust Layers} \\
\multicolumn{4}{c}{}\\ \hline\hline
layer  & $Z$ & $T_b$(K) & $I_{csf}/I_{\rm star}$\\ \hline
a & 32 & $2\cdot 10^8$ & $\sim 2\cdot 10^{-2}$ \\
b & 40 & $3\cdot 10^8$ & $\sim 3\cdot 10^{-3}$  \\
c & 50 & $4\cdot 10^8$ & $\sim 6\cdot 10^{-4}$  \\ \hline
\end{tabular}
\end{table}
This is about the temperature at which crystal lattices usually  become 
brittle and yield to excessive stress by breaking instead of by plastic flow 
(creep) (\cite{rud91}). (A crust's ``Coulomb lattices'' have no natural scale 
so that the ratio of brittle onset temperature to melting temperature should 
not be sensitive to density if the impurity fraction is fixed.) The last 
column is a very rough estimate of the moment of inertia of inter-nuclear
superfluid neutrons in each crustal layer $(I_{csf})$ relative to the
moment of inertia of the star $(I)$. It is extrapolated, very roughly, from 
the nuclear physics calculations of Negele and Vautherin at arbitrarily
selected densities  by assuming 
layer changes occur halfway between those densities at which there is a 
calculation indicating different most stable nuclei. Pinning does not exist in 
all of layer c, and the $I_{csf}$ for layer c only includes the pinning part 
of it. The  $T_b$ are near the estimated deep crust temperatures for the 
$10^3$ year old Crab (and for PSR's 1509-58 and 0540-69).  As a pulsar cools, 
the first crust layer to become brittle ($c$) contains only 
$I_c/(I_a + I_b + I_c) \sim 3\cdot 10^{-2}$ of the total neutron superfluid 
within the brittle crust of older colder pulsars (e.g., Vela).
Because the Crab pulsar would plausibly be just such a pulsar, i.e. one 
with a partly brittle crust, its largest glitches could be smaller by just 
this $3\cdot 10^{-2}$ ratio.  PSR 1509-58 and 0540-69 crusts could be 
sufficiently warm that their crusts are nowhere brittle enough for glitches. 
[Since the supernova remnant around PSR 1509-58 has an age of 20,000 years,
much longer than the pulsar's spin-down age, it has been suggested that the 
pulsar might have been born with a smaller magnetic field 20,000 years ago and 
became a pulsar only about $10^3$ years ago when its magnetic field grew to 
sufficient strength (Blandford, Applegate and Herquist 1983). However, if this 
is the case,
this pulsar should have a much stronger glitch activity. The fact that this
pulsar has never been observed to glitch (Kaspi, et al. 1994) is  strong
support for the presumption that its spin down age is near its true age.]

\item[d)] {\it In addition to giant Vela-like
glitches the much weaker family of Crab-like glitches, is also often observed
in Vela-like and older pulsars (\cite{cor88}). The spread in observed 
$\Delta \Omega/\Omega$ within a family is generally
less than the separation between families.}  As a pulsar cools, crust
magnetic stress  from the pull of spin-down induced flux tube motion in the
core is first relieved by plastic flow (PSRs 1509-58 and 0540-69).  At this 
stage
there is no crust cracking and thus no glitching.  In the slightly cooler
Crab, crust layer $c$ has become brittle and glitching begins in that
layer.  After $10^4$ yr the crust is cool enough that all three layers,
$a$, $b$, and $c$, are brittle and we can now recognize several glitch
families with relative magnitudes for $\Delta \Omega/\Omega$ proportional
to the $I_a$, $I_b$, and $I_c$ of their respective neutron superfluid
moments of inertia ($I_{csf}$ of Table 3).
(This explanation makes the assumption that the shearing 
stress needed to slide two layers with respect to each other, is less than the
stress which would crack  either one.)

\item[e)] {\it Glitch magnitudes, $\Delta \Omega/\Omega$, decrease with 
increasing pulsar period, and glitching essentially ceases at $P=0.7\ {\rm s}$
regardless of pulsar age.} This is shown in Figure\ 8 where the data of 
Fig.\ 5 are replotted as a function of pulsar period. (No account is taken of 
the reduced probability for seeing a glitch in any one pulsar or of the larger 
number of longer period pulsars. The one reported very small pulsar glitch 
(\cite{dow82}) beyond 
this cut off is anomalous in various ways, e.g., in its post-glitch healing.) 
From Equation~(\ref{gact}) drops in $\Delta \Omega/\Omega$ must come from
decreases in $\tau_g/T_s$.  Such decreases are expected when the glitching 
rate is proportional to the speed of the movement through the crust of the 
crust anchored moving core flux tubes.  This tangential speed $(\dot s)$ is 
related to the outward radial velocity of core vortex lines ($v_\perp = v_V$ 
of Equation~\ref{vor}) by
\begin{eqnarray}
\dot s = {v_\perp R\over (R^2 - r_\perp^2)^{1/2}}.
\label{sm}
\end{eqnarray}
Since $\tau_g \sim \Delta s/\dot s$, both $\tau_g$ and $\Delta \Omega/\Omega$ 
(from Equation~(\ref{gact})) approach zero as the core's flux tubes reach the 
core radius at $r_\perp = R$. However, a more quantitative calculation of the 
$r_\perp$ at which glitching should stop must not ignore the finite yield 
strength of the crust. Because of it, crust yielding as well as glitching 
should cease somewhat before $r_\perp = R$ is reached.

The three dashed curves of Figure\ 8 are the predicted $\Delta \Omega/\Omega$ 
from Equation~(\ref{gact}) and Equation~(\ref{sm}) for the three deep 
crust layers of Table 3 with their different $I_{sfn}$. The $r_\perp$ are 
related to pulsar spin-periods by
\begin{eqnarray}
r_\perp = r_\perp (0) \left( {P\over P_0}\right)^{1/2},
\end{eqnarray}
where $r_\perp (0)$ is the distance from the spin-axis of the most important
magnetized surface platelets when the spin period $P=P_0$.  The plotted curves
are for $r_\perp = r_\perp (0) = 0.4 R$ when $P=P_0 = 0.1$~s; $P_0$ is the
spin-period of the Vela pulsar family where $v_\Phi \sim v_V$ is finally
achieved and $r_\perp (0)$ is taken as a plausible estimate. (An $r_\perp (0)$ 
of order half $R$, corresponds to $P\sim 0.5$~s
for canonical large glitch cessation.) The magnitude of
the giant glitches in Vela is determined by using the assumed pulsar and 
glitch independent $\Delta s \sim 2\cdot 10^2\
{\rm cm}$ crust displacement in Crab glitches together with the (calculated)
ratio of crust superfluid moment of inertia to $I_* \sim I$. The smaller glitch
magnitudes are then fixed by the relative moments $I_{a, b, c}$.  The fits
of the model curves in Figure\ 8 seem suggestive of present glitch data. 
\item[f)] {\it Crab glitches occur at intervals 
larger than those between Vela glitches (3 {\rm years}).} Most models
predict (in agreement with observations of  other glitching
pulsars) that the glitching rate is roughly proportional to a pulsar's
spin-down rate.  This would imply that the Crab should glitch at
almost 10 times the rate for Vela.  However in the model of Section 2, the 
glitch rate determined by core flux tube movement, is proportional only to 
the core flux array expansion velocity.  It will no longer be proportional to 
the spin-down rate when superfluid neutron vortices cut through core flux 
tubes as is expected to be the case for the Crab pulsar (cf. Figure\ 1). 
Rather
\begin{eqnarray}
	\tau_g \sim {4T_s\Delta s\over R}\cdot 
	{|{\bf v}_V |\over |{\bf v}_\Phi|}.
\end{eqnarray}
With ${|{\bf v}_\Phi | / |{\bf v}_V|}\sim 0.2$ for the Crab pulsar and $\sim 
0.8$ for the Vela pulsar so that Equation~(\ref{index2}) gives the observed
spin-down indices, the predicted $\tau_g(Crab)\sim 0.4\tau_g(Vela)$. This
only partly accounts for the long $\tau_g(Crab)$. Another contribution to
increasing it might come from some plastic flow to release stress 
in the mainly brittle layer(c). It 
thus appears that there are two separate reasons for the greatly diminished 
glitch activity of the Crab pulsar family, a restricted (or absent) brittle
layer which leads to very small $\Delta\Omega/\Omega$, and a cutting through 
of flux tubes by vortex lines which extends $\tau_g$.
	\item[g)] {\it At least one Crab pulsar glitch has a resolvable 
initial rise in spin-rate (Lyne, Smith and Pritchard 1992, 1993).} After any 
sudden motion of the crust there can be some glitch-like spin-up even in the
absence of any spin-down of crustal neutron superfluid. The positions of 
vortices in the expanding core vortex array are determined by a balance
between the Magnus forces which push the vortices outward and the $10^{15}$
flux tubes per vortex line which encompass each of them and restrain their
outward movement. These flux tubes are anchored by the quasi-rigid highly 
conducting crust. Wherever that crust breaks to relax some of the resulting
stress, the restraining forces on the vortices are diminished and the vortices
may move outward to new positions. How quickly they will do this is (cf. 
Section 2) still unclear and may differ greatly among the superfluid regions.
When the new steady state is finally accomplished there is an increase in 
$\Omega$, the spin of the rest of the star, of roughly
\begin{eqnarray}
	{\Delta\Omega\over\Omega}\sim {\Sigma_{max} I_n^\prime\over 
I\rho_nR^2\Omega^2}\left({l\over R}\right)\left({\Delta s\over R }\right)
\end{eqnarray}
where $\Sigma_{max}$ is the yield stress of crustal matter, $l$ is the crust
thickness, $\Delta s$ is the crust shift in a cracking event (Section 4a ),
and $I_n^\prime$ is the moment of inertia of those core neutrons whose 
spin-down decrement is fast enough to contribute to a glitch observation.
For a typically assumed $\Sigma_{max}\sim 10^{26} 
dyne\ cm^{-2}$ (corresponding to a yield strain $\sim 3\cdot 10^{-4}$), and 
$\Delta s\sim 10^2 cm$ from Section 4a,
\begin{eqnarray}
{\Delta\Omega\over\Omega}\sim {10^{-9}I_n^\prime\over\Omega_2^2I}\ .
\end{eqnarray}
This is too small and has the wrong $\Omega$ dependence to be a significant 
addition to the $\Delta\Omega/\Omega$ of giant glitches, but it may be 
significant for the Crab-like glitch family. It would differ in its initial 
time-dependence from that expected from sudden crustal vortex unpinning:
instead of  an initial (still unresolved ) spin-down as angular momentum is
transferred to core neutrons there would be an initial spin-up as angular 
momentum flows in the opposite directions. This may be suggestive of the 
Crab 1989 glitch but more observations and analyses of the beginning of a 
Crab-like glitch are needed.
\end{description}

\section{Problems}
In this section we discuss  special problems associated with the proposed
model which need further investigation. The first is that the 
total heat generation predicted by the simplified version of the model seems 
too large compared
to the upper bound to it from x-ray observations; the second  is that the
time scale for angular momentum sharing between neutron star-crust and some of
its core neutrons
given by the model seems very much longer than the conventional irresolvably 
short one used in glitch analyses (e.g. \cite{alp88}).
\subsection{Heat generation during neutron star spin-down}
To move outward during spin-down, core vortex lines must either push flux tubes
through the core $e - p$ sea or cut through them. Either would generate heat 
which must be compared to bounds on it from thermal X-ray 
observations of pulsars. When there is no flux-tube cutting and all flux tubes 
are pushed through a core's stationary electron-proton sea, the  
heat production rate would be
\begin{eqnarray}
        \dot Q = \int {\bf\it F}\cdot {\bf v}_\Phi d^3r\sim {\pi\sigma B^2R^5
\over 30c^2T_s^2}\sim 10^{35} \left({B\over 10^{12}G}\right)
\left({R\over 10^6 cm}\right)^5\left({10^4yr\over T_s}\right)^2 erg\ s^{-1}. 
\label{heat}
\end{eqnarray}
But soft X-ray observations of Vela seem to give a bound of $\dot Q\simeq 
10^{33} erg\ s^{-1}$ (\cite{oge93}). This large discrepancy suggests that 
understanding how 
moving core vortex lines move with, or through, the extraordinarily dense flux 
tube array in which they are embedded, without an
unacceptably large $\dot Q$, may be an important question for almost all 
spin-down
models of strongly magnetized pulsars. Below we list various 
possibilities for resolving this problem while still preserving essential 
features of the model proposed in Section 2.
\begin{description}
\item[a)] A most obvious failure of the idealized model is its 
(obviously false) assumption that the core magnetic field of a pulsar can be
approximated as one with enough axial symmetry around ${\bf \Omega}$ so that
outward moving flux tubes must always move through the electron-proton sea in 
which they are embedded. However, this is probably not at all the case in 
regions with inhomogeneously distributed strong core magnetic flux densities. 
Magnetic flux tube, 
vortex lines and e - p plasma might all move together where $n_\Phi$ is very
large without heat 
generation. In that case the integration
volume of Equation~\ref{heat} and the relevant $B^2$ could be much smaller.
\item[b)] In Equation~(\ref{heat}) it has been assumed that vortices are
moving together with flux tubes everywhere in the core. This might not hold 
for the Vela pulsar. If the critical radius of Equation~\ref{r_cr} is only, 
say,  
about one-third of the radius of the Vela pulsar core, the average velocity 
of flux tubes would be roughly three times smaller than that of vortices and 
the total heat generated could be almost an order of magnitude smaller.
\item[c)] A key assumption of the analysis of flux tube drag in being pushed
through the e - p sea plasma is
that magnetic flux tubes are relatively uniformly distributed at least
on the microscopic level. If this is not
the case and some clumping instabilities among flux tubes develops during spin
down, the drag force on the moving flux tubes could be much smaller and
thus  give smaller heat generation. Flux tubes may tend to clump
around the moving vortex lines (about $10^{-2}$ cm away from each other) while 
e - p backflow occurs in between where there are almost no flux tubes. As in
a) a relative motion between flux tubes and the electron-proton sea 
could be restricted to very weak B-field regions.
\item[d)] A type I superconductor might be formed by protons in most of a
neutron star core. From an estimate of the core proton gap energy of 
$\Delta\sim 1MeV$,
it had been argued (e.g. \cite{bay69a}) that core protons form a type II 
superconductor. However a subsequent calculation (\cite{wam90}) which took 
account
of the nuclear interaction between protons and neutrons gave a much smaller 
gap energy ($\Delta\sim 0.2 - 0.3 MeV$). It is then somewhat
less clear whether 
the core protons form a type II or a type I superconductor. For a stiff 
equation of state part of the core protons may well 
form a type I superconductor, while for a soft equation of state it is probable
that only the type II superconductor exists in the core protons of a
neutron star.
Evidence supporting an intermediately stiff or a stiff equation
of state (\cite{lin92}) suggests protons might indeed form a
type I superconductor in part of the  core. There, magnetic field would
be in a mixed state in which B becomes large enough ($\sim 10^{15}Gauss$)
to quench superconductivity in some small slab-like regions, and essentially 
vanishes  in between them. The typical size of such field-free regions is about
$(L\xi)^{1/2}B_c/B\sim 1cm$ with $L\sim 10^6 cm$ the assumed scale size 
of the type I superconducting region. The type I region can also
influence  flux tubes in type II region to bunch together on a similar 1cm 
scale. This could significantly reduce drag forces and thus $\dot Q$. 
\item[e)] Some $\dot Q$ might escape from the 
star's near environment as hard unobserved UV that the soft X-ray observation
bound for $\dot Q$ is significantly exceeded. In young $\gamma$-ray pulsars 
such as Vela there are plausible mechanisms for the generation of $e^\pm$ 
clouds all around the near environment of the pulsar. 
Because of the huge $e^+/e^-$ 
cyclotron resonant scattering of X-ray photons of energy $e\hbar B/mc$, an 
energy which extends from 20KeV to 20 eV within 10 stellar radii, this $e^\pm$
atmosphere would be optically thick to thermal X-rays for plausible $e^\pm$ 
densities (\cite{zhu97}). Much of the emitted soft X-rays might then be 
degraded to hard UV before escaping through this magnetized lepton ``blanket''.
\end{description}

Among all of the above possibilities {\bf a)} would appear most likely 
to be important, i.e. a fundamental inadequacy of the idealized model for
core flux tube motion (especially in layers not adjacent to the crust core 
interface).
\subsection{The initial glitch time scale}
The time scale ($\tau_{spin-up}$) for a suddenly spun-up crust, in a glitch,
sharing its tiny angular momentum jump with the core's much heavier 
superfluid neutrons is usually taken to be unobservably short (\cite{alp93}). 
Because it is not resolved in Vela pulsar this time scale is presumed to be 
less than $10^2 s$ (\cite{mcc90}; \cite{fla90}).  The value estimated from our 
proposed model or any model which involves flux-tube drag or cutting-through 
can give a very different result. Because of the drag on the $10^{14}$ flux 
tubes that must be carried inward or cut through by each of Vela's core vortex 
lines to accomplish a small rapid increase in core neutron angular rotation 
speed, the response of these superfluid neutrons  may be very sluggish. 

For Vela's core's superfluid neutrons very quickly to share in the angular 
momentum given up by crustal superfluid neutrons in a glitch, the core 
neutrons' vortices must move inward about 1cm in less than $10^2$s. Before 
this occurs the core vortex array first increases its rotational speed in
response to the sudden spin-up of the core's flux tubes with which these 
vortices interact. This causes an incremental inward push (Magnus force) on
the core neutron vortices. This force density
\begin{eqnarray}
	\delta F\sim n_r{\pi\hbar\over m_n}\rho_n\delta\Omega R = 
\delta\Omega \Omega\rho_n R
\end{eqnarray}
where $\delta\Omega\sim 10^{-4}\Omega$ is the initial (unresolved) giant
glitch spin-up before there is any transfer of angular momentum to core
superfluid neutrons. If the subsequent inward vortex motion involves 
pushing flux tubes through the electron proton sea, Equation~\ref{vphi} gives
a maximum inward flux tube speed
\begin{eqnarray}
 \delta v_\Phi\sim {\delta\Omega\Omega\rho_n R c^2\over \sigma n_\Phi\Phi_0^2}
\sim 10^{-11} cm\ s^{-1}\ .
\end{eqnarray}
To move inward by 1cm would then take
\begin{eqnarray}
	\tau_{spin-up}^\prime\sim 10^{11}s\sim T_s
\end{eqnarray}
Where flux tube cut-through by moving vortices occurs first the time scale
$\tau_{spin-up}^\prime \gg 10^2s$ for $B\sim 10^{12}G$ (\cite{din93}).
Almost all of the possibilities in Section 5.1
for reducing $\dot Q$ would also reduce 
$\tau_{spin-up}$, but for some, or perhaps all, core neutrons the needed 
reduction seems so large  that it is hard
to see how $\tau_{spin-up}$ can become unobservably short for all of the
core neutron superfluid. 
One possibility for resolving this problem may be to accept the
model result that where vortices must push flux tubes through the 
electron-proton sea or cut through them, $\tau_{spin-up}$ is unresolved 
because it is too long,
i.e. far longer than the interval between glitches ($\tau_g$). With the 
possible resolution suggested in Section 5.1a), those vortex lines whose 
surrounding flux tubes move with their embedding e - p sea may quickly adjust
($\tau_{spin-up}< 10^2 s$) and also generate little $\dot Q$, while only 
a very small minority of
vortex lines with the flux tubes they carry actually move through their local 
charged sea.
If this is the case, although the
$I_*$ of Equation~(\ref{gact}) would  not include all core superfluid neutrons,
it still might be nearly the entire I of the star. This would also
be the case if the core is mainly a K-condensate or quark matter, 
superconductors with no purely neutral superfluids to be spun-up in a glitch. 
( The charged ones are easily
spun-up by any magnetic field which couples them to the crust.) It should
be noted that a large reduction of $\tau_{spin-up}$ for some parts of the
core neutron superfluid could put the time scale in the range where it should
contribute to glitch ``healing'' analyses.

\acknowledgments
It is a pleasure to thank A. Alpar, K.S Cheng, P. Goldreich, F. Graham-Smith,
A. Lyne, and D. Pines  for informative conversations.
This work was supported in part by NASA grants NAG 5-2016. 

\appendix
\section{Superfluid-superconductor interactions}
Because magnetic field inside neutron stars are usually not aligned along
the spin axis when neutron stars spin-down (-up) the outward (inward) moving 
superfluid neutron vortices run into proton flux tubes. The interaction 
between superfluid neutron vortices and proton superconductor magnetic flux 
tubes as they try to cross through each other can thus play an important part 
in determining the motion of both vortices and flux tubes. Srinivasan et 
al. (1990) 
proposed that the proton density perturbation in the center of a flux tube
would give rise to an interaction energy per intersection 
\begin{eqnarray}
	E_{int}\sim n_n{\Delta_p^2\over E_{F_p}^2}{\Delta_n^2\over E_{F_n}} (
	\xi_n^2\xi_p)\simeq 0.1 MeV \ , \label{intforce}
\end{eqnarray}
where $\xi_{n,p}$ are the neutron, proton BCS correlation lengths, 
$\Delta_{p,n}$ are
the respective gap energies,  $E_{F_{p,n}}$ the Fermi energies and $n_n$ the
neutron number density.
An even more important contribution to the interaction energy comes from 
the magnetic interaction between neutron vortex lines and proton flux tubes
and from the velocity dependence of the nuclear interaction between the
neutrons in a vortex and the protons in a flux tube, which is also the 
ultimate cause of the
neutron vortex line flux. Both can be taken into account using an 
effective Ginzburg-Laudau (GL) free energy ($f_{GL}$) for an interacting 
mixture of superfluid neutrons and superconducting 
neutrons (\cite{alp84b})
\begin{eqnarray}
	f_{GL} = f_u + {1\over 2}\rho_s^{pp}v_p^2 + {1\over 2}\rho_s^{nn}v_n^2
+ \rho_s^{pn}{\bf v}_p {\bf \cdot }{\bf v}_n + {B^2\over 8\pi},
\label{energy}
\end{eqnarray}
where $f_u$ is the condensation energy density, $\rho_s^{pp}$ and $\rho_s^{nn}$
are the ``bare'' densities of superconducting protons and superfluid neutrons
respectively, $\rho_s^{pn}$ is the coupling density, and ${\bf v}_p$  and 
${\bf v}_n$ are the superfluid velocities defined by
\begin{eqnarray}
	{\bf v}_p & = &{\hbar\over 2m_p}{\bf\nabla }\chi_p - 
	{e\over m_pc}{\bf A},
\label{vp} \\
	{\bf v}_n & = &{\hbar\over 2m_n}{\bf\nabla }\chi_n.
\label{vn}
\end{eqnarray}
The superfluid electric current is 
\begin{eqnarray}
	{\bf j}_s \equiv {c\over 4\pi}({\bf\nabla}\times {\bf B})= {e\over m_p}
[\rho_s^{pp}{\bf v}_p + \rho_s^{pn}{\bf v}_n]\ .
\label{js}
\end{eqnarray}
From Equations~(\ref{js}) and (\ref{vp}), (\ref{vn}) we obtain  
London's equation 
\begin{eqnarray}
	\nabla^2{\bf A} - {{\bf A}\over\Lambda_*^2}= -{2\pi e\hbar\over m_p^2c}
	[\rho_s^{pp}{\bf\nabla }\chi_p + 
	\rho_s^{pn}{m_p\over m_n}{\bf\nabla }\chi_n]
\end{eqnarray}
with $\Lambda_* = (m_p^2c^2/4\pi e^2\rho_s^{pp})^{1/2}$ the effective London
penetration depth. For a pure proton flux tube with $\nabla\chi_p = {\hat{\phi}\over r}$ and $\nabla\chi_n = 0$, the above equations give
\begin{eqnarray}
      {\bf v}_n & = & 0;\\
      {\bf v}_p & = & {m_p\over\rho_s^{pp}e}{c\over 8\pi\Lambda_*}{\Phi_0\over\pi\Lambda_*^2} K_1\left({r\over\Lambda_*}\right);\\
       B & = &{\Phi_0\over 2\pi\Lambda_*^2} K_0\left({r\over\Lambda_*}\right),
\end{eqnarray}
with $\Phi_0 = \pi\hbar c/e$ the flux quantum and $K_0$ and $K_1$
Bessel functions of order zero and one with imaginary argument.
The solutions for a pure neutron vortex line with 
$\nabla\chi_p = 0$ and $\nabla\chi_p = {\hat{\phi}\over r}$ or a 
superposition of a neutron vortex line and a proton flux tube 
 with $\nabla\chi_p = {\hat{\phi}\over r}$ and $\nabla\chi_p = {\hat{\phi}\over r}$ can be obtained similarly.
\begin{eqnarray}
      {\bf v}_n & = &{\hbar\over 2m_n}{\hat{\phi}\over r},\\
      {\bf v}_p & = &{m_p\over\rho_s^{pp}e}{c\over 8\pi\Lambda_*}{\Phi_*\over\pi\Lambda_*^2} K_1\left({r\over\Lambda_*}\right) - {\hbar\rho_s^{pn}\over 2m_n\rho_s^{pp}}{\hat{\phi}\over r},\\
      B & = &{\Phi_*\over 2\pi\Lambda_*^2} K_0\left({r\over\Lambda_*}\right).
\end{eqnarray}
with $\Phi_*$ the total flux in a single flux tube. For an isolated neutron 
vortex line $\Phi_* =\Phi_0(m_p\rho_s^{pn}/ m_n\rho_s^{pp})$. For a 
superimposed vortex line and flux tube 
$\Phi_* = \Phi_0[ 1+{m_p\rho_s^{pn}/m_n\rho_s^{pp}}]$.

The energy for each case can be estimated from 
Equation~(\ref{energy}). The extra energy (per unit length) of the 
superposition of a flux
tube and a vortex line relative to a distantly separated flux tube and a
vortex line is
\begin{eqnarray}
 E \simeq {\pi\over 8}\left({\Phi_0\over\pi\Lambda_*^2}\right)^2\Lambda_*^2{m_p\over m_n}{\rho_s^{pn}\over\rho_s^{pp}} \ln\left({\Lambda_*\over\xi}\right) \ .
\end{eqnarray}
There are many more flux tubes than vortices. We assume that just
before cutting through the typical distance between two consecutive flux tubes 
pushed by the same moving vortex is about $\Lambda_*$, i.e. flux tubes are
swept up by a moving vortex but not cut through. The magnetic repulsion 
between flux tubes limits their density. This repulsion is not effective until 
the inter-flux tube separation approaches $\Lambda_*$. Then the 
maximum force density on a flux tube array would  be roughly estimated as
$E/\Lambda_*$ or
\begin{eqnarray}
	F_{max} \simeq {\pi n_V\over 8}
\left({\Phi_0\over\pi\Lambda_*^2}\right)^2\Lambda_*{m_p\over m_n}{\rho_s^{pn}\over\rho_s^{pp}} \ln\left({\Lambda_*\over\xi}\right) 
= {\pi n_V\over 8}B_VB_\Phi\lambda_*\ln\left({\Lambda_*\over\xi}\right)\ ,
\end{eqnarray}
with $n_V$ the number density of vortex lines,
$B_\Phi = \Phi_0 /\pi\Lambda_*^2$ the characteristic magnetic field in the
cores of flux tubes and $\ B_V = (\Phi_0 /\pi\Lambda_*^2)
(m_p \rho_s^{pn}/ m_n\rho_s^{pp})$ the field within the cores of neutron
vortex lines which are embedded in the stellar core's superconducting
proton sea.

\newpage
\centerline{\large\bf Figure Captions}
\figcaption[fig.1]{ Radial vortex line speed $v_V$ and induced flux tube 
radial speed $v_\Phi$ vs. the radial distance to the spin-axis ($r_\bot$). 
For $r_\bot < r_c,\ v_\Phi = v_V$; for $r_\bot > r_c, \ v_\Phi < v_V$. It is 
not yet known how far $v_\Phi$ drops below $v_V$ when $r_\bot > r_c$ and two
linear possibilities are indicated.  
\label{fig1}}
\figcaption[fig.2]{ Model evolution of magnetic dipole fields of radiopulsars.
Star-like designations indicate radio pulsars found in SNRs.  In the model
solitary 
spinning-down radio pulsars follow the path (a--b--c). The path (a--b) 
corresponds to
the first and second stages discussed in Section 3.  Spin-down follows the path
(b--c--d) when field-pulled parts of the crust move toward the spin-equator 
where reconnection
can begin after core flux expulsion. The region (d) would not be reached by a
solitary pulsar, but may be by some neutron stars in binaries. Further 
spin-down
beyond (d) would not be effective in reducing $B$ because the crust would no
longer be stressed above its yield strength. (Subsequent accretion induced
spin-up could return the neutron star to (c) if the magnetic field
configuration mainly connects the two spin hemispheres.) 
\label{fig2}}
\figcaption[fig.3]{ Model for movement of a single magnetic flux tube in a 
spinning-down 
neutron star core. (a) Side view of initial flux tube path (thicker line). In 
the crust and beyond, the magnetic field is not confined to quantized 
flux tubes. Neutron superfluid vortex lines are indicated as unfilled tubes.  
Because the core field would be expected to have had toroidal as well as 
poloidal components before the superconducting transition, the flux tube path 
is probably quite tortured while the vortex array is quasi-uniform. (b) Top 
view of (a) from along the spin axis direction. (c) Top view of the flux 
tubes in the equatorial zone after  long spin-down. A conducting crust 
platelet moves with the flux tube capitals,  pushed beyond the crust's yield 
strength in part by the crust's own pinned vortex lines and , crucially, by 
the pull of core flux tubes. As  core neutron vortex
motion moves an entrained flux tube, that tube is ultimately pushed into the
crust core boundary for almost any initial flux tube configuration.
\label{fig3}}
\figcaption[fig.4]{ Movement of magnetized patches (``platelets'')
on the surface of a 
spinning-down pulsar: a) initial surface magnetic field configuration; b)
after substantial spin-down the main (most strongly magnetized) patches have 
reached the spin-equatorial 
zone where reconnection can occur; c) remaining magnetized patches after 
reconnection. The magnitude of ${\bf B}$ at the patch N$^\prime$ 
remains about the same as its initial one in a), but the dipole moment 
(${\bf \mu}$) has become much smaller and its orientation is changed.
\label{fig4}}
\figcaption[fig.5]{Fractional jumps in pulsar spin-rate $(\Omega)$ in glitches 
as a function of the spin-down age $(P/2\dot P)$ of the glitching
radio-pulsars (Lyne et al.\ 1995).
\label{fig5}}
\figcaption[fig.6]{Pulsar glitch activity vs. pulsar spin age  from 
Lyne et al. (1995). The dots are PSRs 0833, 1338, 1737, 1823, 1758.  
The diagonal line is the glitch activity from Equation (18) with 
$I_s/I_* = 1.5* 10^{-2}$.
\label{fig6}}
\figcaption[fig.7]{The rotation frequency of the Crab pulsar over a 23-year 
period after subtracting an extrapolation from the first few years of data (
Lyne et al. 1992).
\label{fig7}}
\figcaption[fig.8]{Observed glitch magnitudes (Lyne et. al.\ 1995) vs. pulsar 
period.
\label{fig8}}

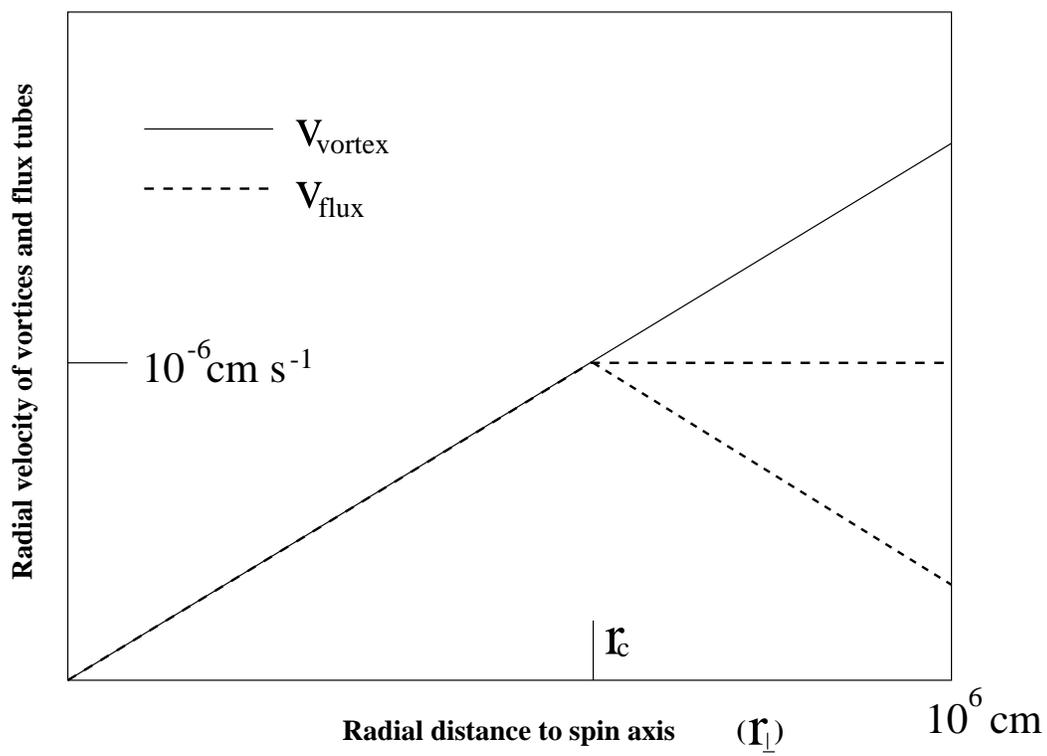
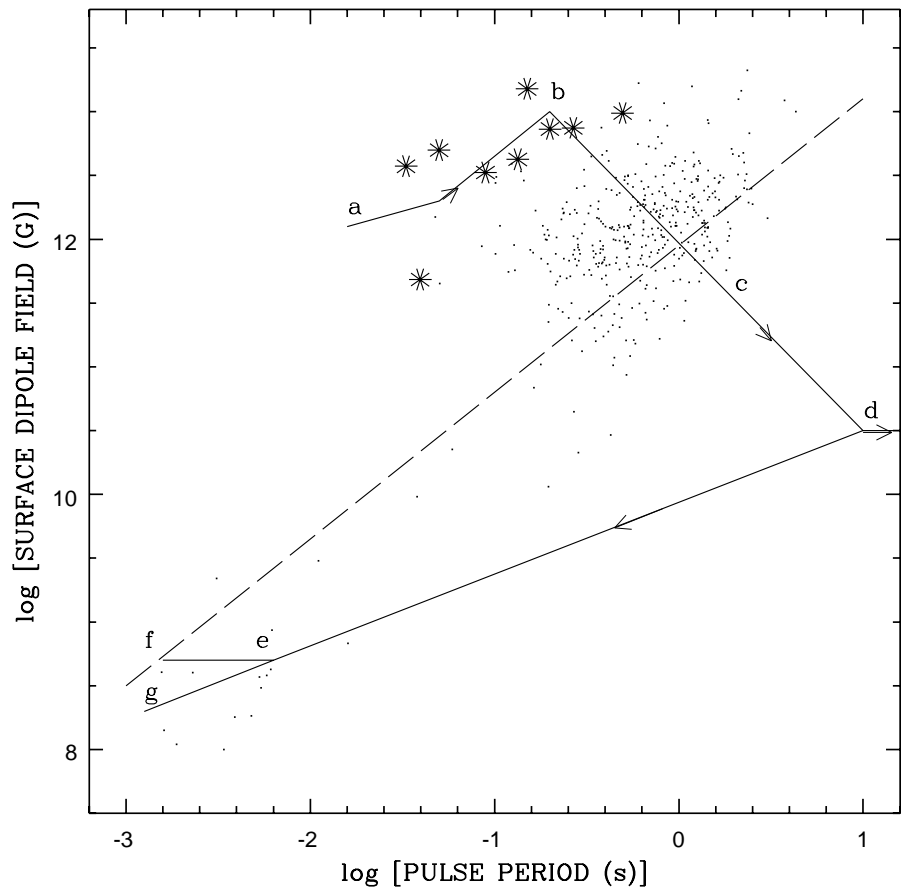
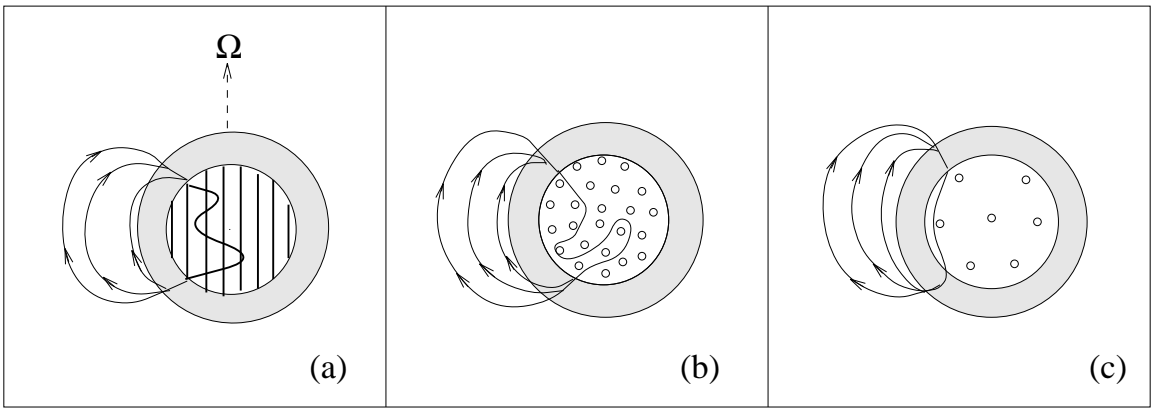
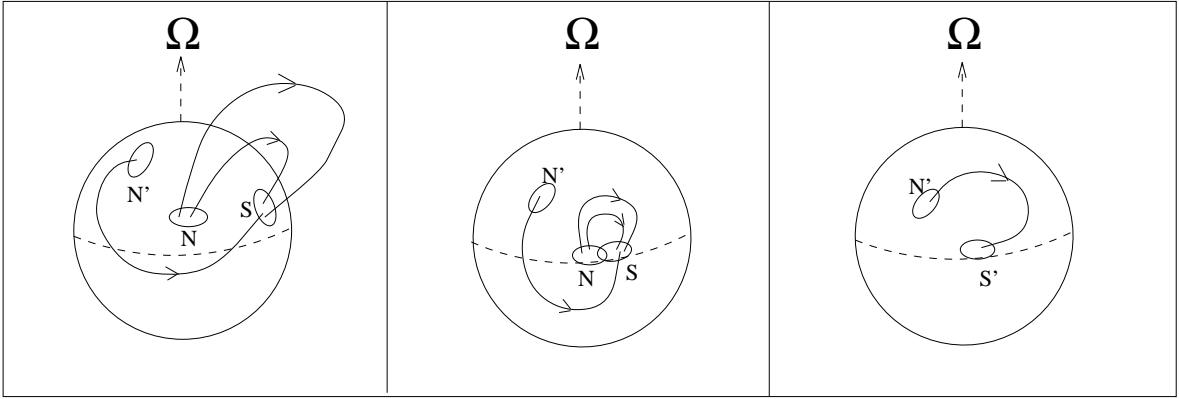
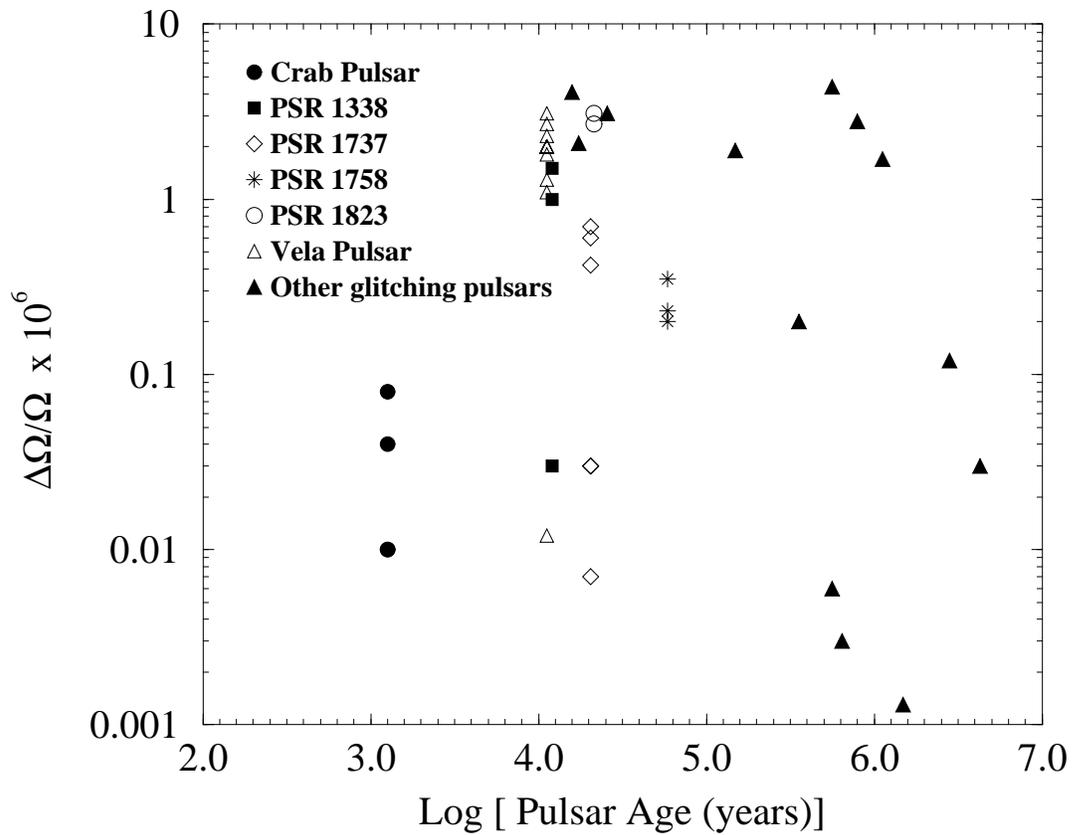
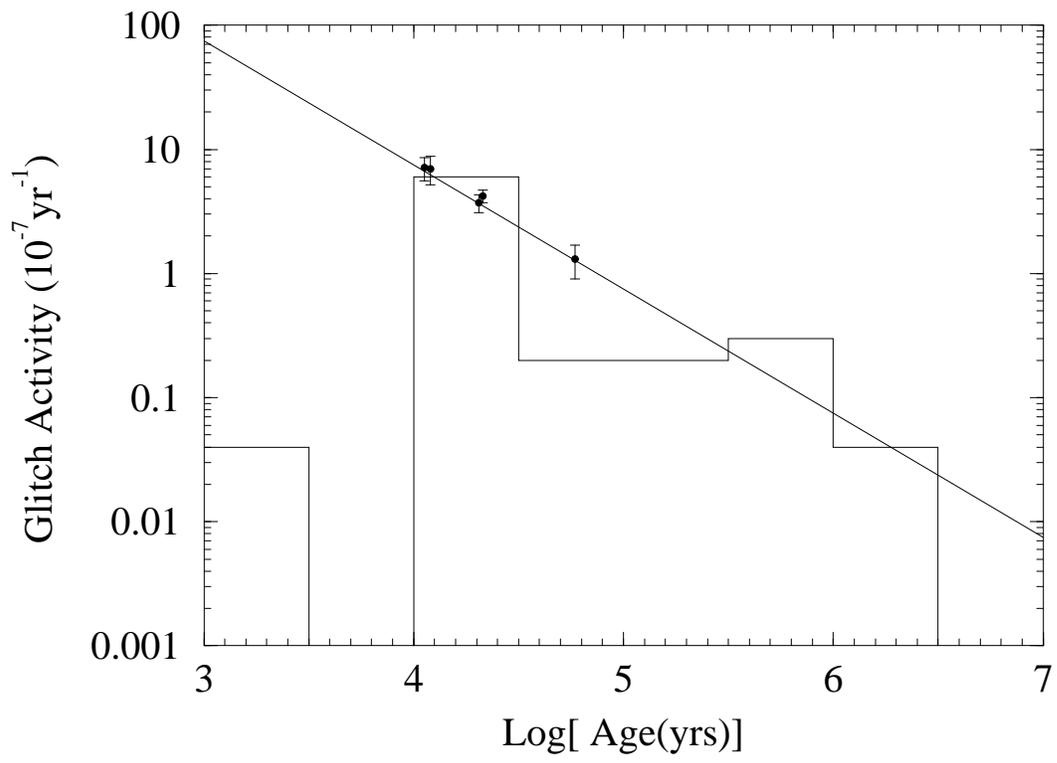
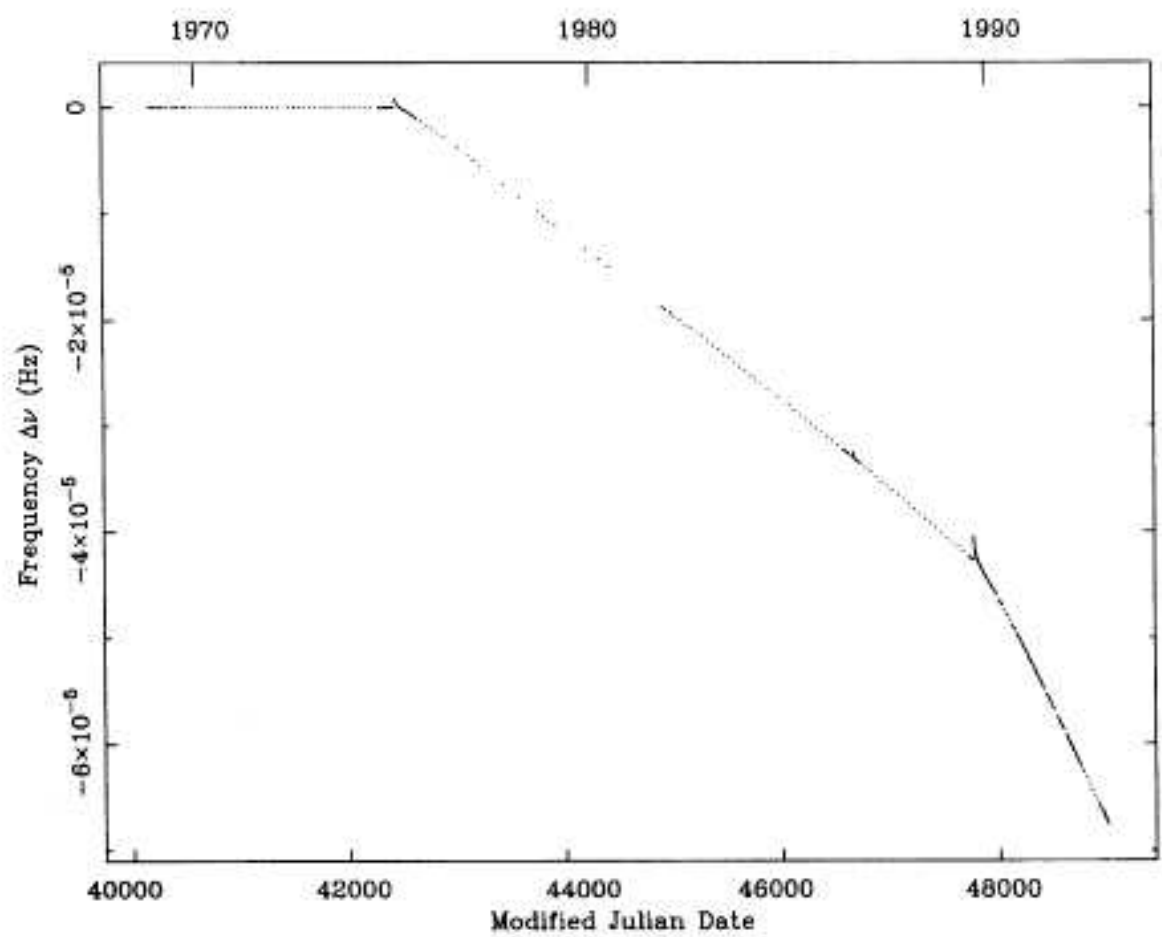
\end{document}